\documentclass[aps,prl,twocolumn,letterpaper,superscriptaddress,showpacs]{revtex4}
\usepackage{keyval}%
\usepackage{graphicx}
\usepackage{dcolumn}
\usepackage{bm}
\usepackage{color}

\setkeys{Gin}{width=0.8\columnwidth,clip=true}

\newcommand{\ignore}[1]{}

\begin{document}
\title{Dynamical Linear Response of TDDFT with LDA+\textit{U} Functional:\\
strongly hybridized Frenkel excitons in NiO}
\author{Chi-Cheng Lee}
\affiliation{Condensed Matter Physics and Materials Science
Department, Brookhaven National Laboratory, Upton, New York 11973,
USA}%
\affiliation{Department of Physics, Tamkang University, Tamsui, 
Taiwan 25137, Republic of China}
\author{H. C. Hsueh}
\affiliation{Department of Physics, Tamkang University, Tamsui, 
Taiwan 25137, Republic of China}
\author{Wei Ku}
\affiliation{Condensed Matter Physics and Materials Science
Department, Brookhaven National Laboratory, Upton, New York 11973,
USA}%
\affiliation{Physics Department, State University of New York, Stony
Brook, New York 11790, USA}

\date{\today}

\begin{abstract}
Within the framework of time-dependent density-functional theory (TDDFT),
we derive the dynamical linear response of LDA+\textit{U} functional
and benchmark it on NiO, a prototypical Mott insulator.
Formulated using real-space Wannier functions, our computationally inexpensive framework gives
detailed insights into the formation of tightly bound Frenkel excitons with reasonable accuracy.
Specifically, a strong hybridization of multiple excitons is found to significantly modify
the exciton properties.
Furthermore, our study exposes a significant generic limitation of adiabatic approximation 
in TDDFT with hybrid functionals and in existing Bethe-Salpeter-equation approaches,
advocating the necessity of strongly energy-dependent kernels in future development.
\end{abstract}


\maketitle

Computing many-body excitations of weakly to strongly interacting 
materials continues to be a major challenge for first-principles studies~\cite{Onida}. 
Following the great achievement of Density Functional Theory (DFT) in ground-state calculations
using the local density approximation (LDA)~\cite{Hohenberg,Kohn},
time-dependent density functional theory (TDDFT)~\cite{Runge,Leeuwen} has promised to be
an affordable and accurate theoretical framework to study the dynamical linear response
of weakly interacting materials ranging from finite~\cite{Petersilka,Vasiliev,Yabana}
to extended systems~\cite{Onida,Needs,Ku,Ku3}.
While TDDFT is formally exact in describing the time-dependent density, in practice the lack
of knowledge of the proper functional form of the action functional highlights
the immaturity of its full potential in studying dynamics of real materials.
For example, there isn't yet an affordable first-principles local functional that includes
proper particle-hole interactions at the two-particle level to allow the exciton formation,
leaving the computationally expensive perturbation theory~\cite{Onida,Tokatly}
(assuming not too strong interactions) or the highly parameterized cluster model~\cite{Haverkort}
the only option.
Due to this well-known limitation, the applicability of TDDFT (with existing approximation)
to strongly correlated materials remains basically unexplored.



Recently, a significant improvement in the ground state calculations of strongly interacting
Mott insulators is made via the new LDA+\textit{U}
functional~\cite{Anisimov2,Shick}, in which strong intra-atomic Coulomb interactions are
introduced at the screened Hartree-Fock level.
Similar to the hybrid functionals~\cite{Tran}, the inclusion of non-local exchange
allows opening of the Mott gap and gives also reasonable quasi-particle excitation energies.
(Here we do not limit TDDFT or DFT to the Kohn-Sham framework, and we consider LDA+$U$ and
other hybrid functionals as implicit functionals of density within DFT.)
This physically motivated approach has led to important understanding 
of orbital and charge ordered systems~\cite{Leonov,Jeng,Yin,Dmitri}.
On the other hand, to date the study of charge excitation within LDA+\textit{U} scheme has been
very limited and its applicability to strongly interacting Mott insulators remains unclear.
It is thus interesting and timely to explore the dynamical linear response of LDA+\textit{U}
functional within the framework of TDDFT.

In this Letter, we examine the strength and weakness of LDA+\textit{U} in describing charge
excitations of strongly interacting Mott insulators,
by developing diagrammatically the dynamical linear response of LDA+\textit{U} functional
within TDDFT framework (TDLDA+\textit{U}).
The resulting formula in tackling local excitations connects TDLDA+\textit{U} (and
other hybrid functionals) to a Bethe-Salpeter equation (BSE)~\cite{Salpeter} with intra-atomic
Hartree-Fock kernels.
The framework is then implemented on the basis of symmetry-respecting Wannier functions
(WFs)~\cite{Ku2,Yin}, which not only dramatically reduce the computation expense but also facilitate a comprehensive real-space picture of local excitons.
The integrated methodologies are applied to the study of tightly bound $d$-$d$ Frenkel excitons
in NiO, a representative Mott insulator.
Our diagrammatic approach allows a step-by-step elucidation of the effects of different
components of the interaction kernel.
Specifically, multiple tightly bound (by $\sim$6eV) excitons form inside the Mott gap
and are found to hybridize strongly to give reasonable excitation energies and highly
anisotropic \textbf{q}-dependent spectral weights observed recently~\cite{Larson}.
Furthermore, our results illustrate a serious general limitation of adiabatic approximation widely employed in most state-of-the-art TD-hybrid methods~\cite{Kresse} and BSE
approaches~\cite{Onida2,Rohlfing,Benedict}, and advocate the necessity of energy dependence
in future design of approximations within these theoretical frameworks.

\begin{figure}[tbp]
\includegraphics[width=0.70\columnwidth,clip=true,angle=270]{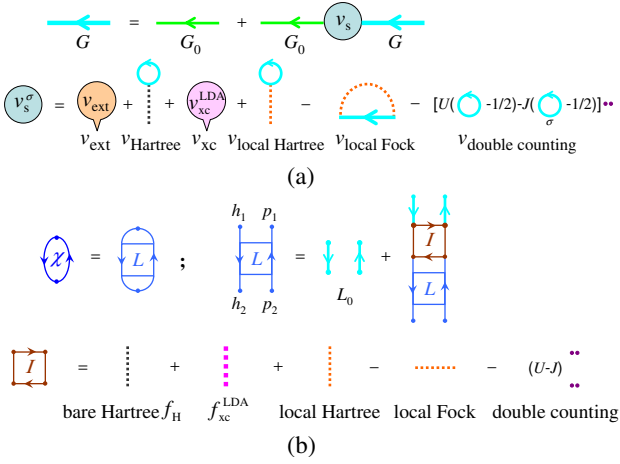}
\caption{\label{fig:fig1} (color online). (a) Dyson equation with LDA+\textit{U} 
potential $v_{s}$. (b) Dynamical linear response function $\chi$, 
correlation function $L$($p_{1}$$h_{2}$;$h_{1}$$p_{2}$), and the corresponding BSE with LDA+\textit{U} kernel.}
\end{figure}

Following the standard procedure~\cite{Baym}, the linear response can be derived from the
equation of motion (Dyson equation) of Kohn-Sham particle shown diagrammatically in
Fig.~\ref{fig:fig1}.
In addition to LDA potentials, site-local (s-local) screened Hartree-Fock interactions 
and the double counting terms among local orbitals are introduced in LDA+\textit{U}.
Taking the derivative of LDA+\textit{U} Green's function with respect to the external potential,
the response function $\chi$,
\begin{equation}
\chi(\textbf{x}_{1}t_{1};\textbf{x}_{2}t_{2})=
\sum_{p_{1}h_{1}p_{2}h_{2}} M^{*\textbf{x}_{1}}_{p_{1},h_{1}}L_{p_{1}h_{2};
h_{1}p_{2}}(t_{1}t_{2};t_{1}t_{2})M^{\textbf{x}_{2}}_{p_{2},h_{2}},
\end{equation}
is formulated by creating s-local particle-hole (\textit{p}-\textit{h}) pairs at position
$\textbf{x}_{2}$ and time $t_{2}$, with probability amplitude
$M^{\textbf{x}_{2}}_{p_{2},h_{2}}\equiv\phi^{*}_{p_{2}}(\textbf{x}_{2})\phi_{h_{2}}
(\textbf{x}_{2})$, followed by the propagation of the \textit{p}-\textit{h} pairs
expressed in terms of the \textit{p}-\textit{h} correlation function $L$, and finally the
annihilation of s-local \textit{p}-\textit{h} pairs at $\textbf{x}_{1}$ and a later time $t_{1}$
with probability amplitude $M^{*\textbf{x}_{1}}_{p_{1},h_{1}}$ as shown in Fig.~\ref{fig:fig1}(b). 
The equation of motion of $L$ (BSE) involves irreducible kernel $I$, including 
bare Hartree ($f_\mathrm{H}$), exchange-correlation $f_\mathrm{xc}$, 
s-local screened Hartree-Fock, and double counting terms. 
The s-local Fock interaction directly provides the 
attraction between \textit{p}-\textit{h} pairs of orbitals that \textit{U} is applied to. 
For the study of local $d$-$d$ excitations, the original $f_{LDA}$ and the double-counting
contributions are meant to counter each other approximately and should be neglected, leaving only
the s-local screened Hartree-Fock contribution.
(Besides, the long-range screening of $f_\mathrm{H}$ is inefficient at short distance.)
This diagrammatic representation of the equation of motion makes it apparent that one is allowed
to visualize the physical effects of interaction kernels by turning them on one after
another, as presented below.

For the study of local excitations, it is most convenient and efficient to employ a real-space
s-local basis~\cite{Soininen}.
 Furthermore, it is advantageous to have orbitals that diagonalize the one-particle density matrix such that they enter only as either pure particle or hole orbitals.
To this end, we implemented the above TDLDA+\textit{U} framework using the
energy-resolved symmetry-respecting Wannier functions~\cite{Ku2,Yin} that are constructed
without mixing the occupied and unoccupied bands~\cite{supplement}.
In this basis, the complicated six-dimensional exciton wave function, $\psi_{i}(\textbf{xx}^\prime)$,
can be decomposed into \textit{small} number of ``bare'' exciton wave functions as direct
products of one particle ($\phi_{p}$) and one hole ($\phi_{h}$) orbitals,
$\psi_{i}(\textbf{x}\textbf{x}^\prime)=\sum_{ph} c^{i}_{ph}\phi_{p}(\textbf{x})\phi_{h}^{*}(\textbf{x}^\prime)$.
This gives an easy visualization of exciton wave functions and a direct computation of various
experimental ``form factor''.
Moreover, the strong binding suppresses significantly the kinetics
of the exciton such that only very few short-range neighbors are necessary in solving
the BSE, significantly reducing the computational expense.
In fact, even a single-site calculation is found quite accurate for NiO~\cite{supplement}.

The LDA+\textit{U} calculation for type-II antiferromagnetic (AFM) NiO (\textit{U}=8eV, \textit{J}=0.95eV) is performed by an all-electron full-potential method using linearized
augmented plane wave basis~\cite{Schwarz}.
It gives the correct ground state of NiO with high-spin configuration,
leaving only the spin-minority channel active for s-local $d$-$d$ charge excitations.
The rhombohedral symmetry, dictated by the AFM order along [111] direction, splits the $d$-Wannier
orbitals into two unoccupied $e_{g}$, two occupied $e^\prime_{g}$, and one occupied $a_{g}$
orbitals.
In the rest of the Letter, we denote
1$\rightarrow$($e_{g1},a_{g}$), 
1$^\prime\rightarrow$($e_{g2},a_{g}$), 
2$\rightarrow$($e_{g1},e_{g1}^\prime$), 
2$^\prime\rightarrow$($e_{g2},e_{g2}^\prime$), 
3$\rightarrow$($e_{g2},e_{g1}^\prime$), and
3$^\prime\rightarrow$($e_{g1},e_{g2}^\prime$) for different \textit{p}-\textit{h} pairs
(\textit{p},\textit{h}) forming bare excitons.
Within this notation, $\chi(\textbf{q},\omega)$ is thus
$\sum_{ij} M^{*\textbf{q}}_{i}L_{ij}(\textbf{q},\omega)M^{\textbf{q}}_{j}$.
Here $M^{\textbf{q}}_{i}\equiv\int{e^{i\textbf{q}\cdot\textbf{x}}M^{\textbf{x}}_{i}d\textbf{x}}$
gives the probability amplitudes in momentum space corresponding to the anisotropic form factor
of inelastic X-ray and electron scattering~\cite{Larson,Muller}.

\begin{figure}[tbp]
\includegraphics[width=0.95\columnwidth,clip=true]{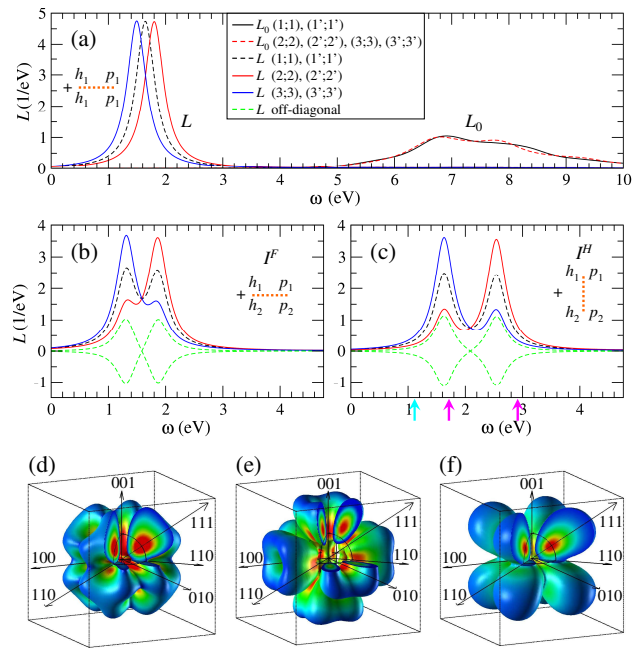}
\caption{\label{fig:fig2} (color online). Imaginary parts of \textit{p}-\textit{h} correlation function
$L_{ij}$ denoted as (i;j) by solving BSE for (a) $L_{0}$: unbound \textit{p}-\textit{h} pair,
and $L$: bare exciton, (b) $L$ with s-local Fock, and (c) $L$ with s-local Hartree-Fock.
Experimental excitation energies are indicated by the arrows.
(d), (e), and (f) illustrate the spectral weight of $L_0$, low-energy exciton in $L$, and
high-energy exciton in $L$, respectively, via 3D isovalue contours in \textbf{q}-space.
The black solid circles indicate \textit{q}=7{\AA}$^{-1}$.
Spectral weight in (e)/(f) resembles very well the experimental ones corresponding to the
blue/red arrows.}
\end{figure}

We now demonstrate in Fig.~\ref{fig:fig2} the physical effects of the above interaction kernels
on forming the low-energy Frenkel excitons in NiO, by switching them on one after another in
the corresponding BSE of TDLDA+\textit{U}.
(A broadening of $\eta$=0.2eV is introduced in calculating $L$ in accordance to the experimental resolution~\cite{Larson}.)
As a reference, the unbound \textit{p}-\textit{h} excitations, $L_0$, containing no interaction
kernel, is shown in Fig.~\ref{fig:fig2}(a).
$L_0$ gives basically the creations of \textit{p}-\textit{h} pairs across sites
(inter-site $d$-$d$ excitations) that contribute at small momentum but are eventually overwhelmed
by the on-site excitations at larger momentum.
Not surprisingly, the excitation energies correspond to the energy difference of LDA+\textit{U}
eigenvalues encapsulating a large Coulomb interaction (of order $U$) due to the addition of a
particle and a hole at different atoms.
In addition, a broad spectrum (5-10 eV) is obtained, reflecting the high mobility of such unbound
\textit{p}-\textit{h} pairs and their strong coupling to the oxygen $p$-orbitals.

For intra-site excitations, however, the interaction kernel can dramatically modify the
excitations.
As shown in Fig.~\ref{fig:fig2}(a), upon switching on the Fock kernel within the \textit{same}
pairs, which encapsulates the strong binding of \textit{p}-\textit{h} pairs,
the excitations collapse into three well-defined doubly-degenerate bare excitons at 1-2eV via
binding energies $\sim$6eV!
This seemingly large binding energy is a necessary consequence of the large s-local Coulomb
repulsion present in NiO and other strongly correlated materials.
Preserving the particle number at each site, the s-local charge excitations should
not be subject to the Hubbard $U$ energy scale of adding or removing one particle.
On the other hand, since one-particle Green's functions used to build $L_0$ includes the energy
scale of $U$, such effects must be countered by a strong attraction of the same scale in any
proper theory of \textit{p}-\textit{h} pairs, as demonstrated here.
Benefiting from the lack of decay process deep inside the Mott gap, these bare excitons have
very long lifetime (negligible line width beyond the experimental resolution).
Due to the well-defined point-group symmetry of these Wannier orbitals (and thus
$M^\textbf{q}_i$),
the resulting bare excitons possess well-defined angular structure in their spectral weight.
As shown in Fig.~\ref{fig:fig2}(d), a prominent anisotropy of $\chi$ is found with the strongest
weight along [111] directions and vanishing weight along [001] directions, in agreement with
the recent experiment~\cite{Larson}.
In addition, the dipole-forbidden nature is revealed clearly by a hollow center at
$q\rightarrow0$.

Next, we activate the scattering process among different \textit{p}-\textit{h} pairs in the
remaining part of the Fock kernel, $I^F$.
These additional couplings between bare excitons introduce an interesting effect of strong
hybridization and split the exciton energies into two sets.
As illustrated in Fig.~\ref{fig:fig2}(b), this strong hybridization results in new large
off-diagonal elements of $L$ in the bare exciton basis, and the double-peak structure in the diagonal elements.
From the conventional eigenvector point of view, this is equivalent to having exciton
eigen-functions of each energy containing superposition of the above bare excitons.
Indeed, the \textbf{q}-dependence of excitons for the low-energy and the high-energy peaks now
shows significant differences, as shown in Fig.~\ref{fig:fig2}(e) and (f).
Along [111] directions, the former has a vanishing weight (new nodal directions), while the
latter shows strong enhancement, reflecting the anti-bonding and bonding nature of the
superposition, respectively.
Such strong hybridization of multi-excitons is absent in weakly correlated systems, but should be
expected from most strongly correlated systems with open shells.

Finally, upon addition of the s-local Hartree kernel, $I^H$, the overall exciton spectrum is
further modified.
As shown in Fig.~\ref{fig:fig2}(c), an overall blue shift of the spectrum is observed, originated
from the screening (weakening) of the exciton binding energy via the diagonal elements of $I^H$.
In addition, without altering the \textbf{q}-dependent spectral weight significantly,
the off-diagonal elements of $I^H$ further split the exciton energies.
Both features are expected from the Kramers-Kronig relation in the typical dynamical screening process.

So far, the above analysis has been performed with Wannier functions associated with only one Ni
site, which makes the proposed framework extremely inexpensive and allows a very clean real-space
physical picture.
More extensive calculations~\cite{supplement} including non-s-local (beyond one Ni site)
effects resulted in a negligible energy reduction ($<$0.1eV) of the overall spectrum,
and, in perfect agreement with the experiment~\cite{Larson}, a negligible dispersion.
Indeed, the Frenkel excitons in NiO are so tightly bound and so heavy to propagate that
the physics can be captured quite completely even with only a single site.
This further advocates strongly our choice of employing Wannier basis for strongly correlated
materials in general~\cite{Ku2,Yin,Larson,Dmitri}.

Overall, compared with the recently measured local $d$-$d$ excitations~\cite{Larson,Muller},
the resulting theoretical spectrum of above TDLDA+\textit{U} performs unexpectedly well for
such a simple approximation.
In great contrast to the qualitative failure of TDLDA~\cite{Ullrich} or the RPA response of LDA+\textit{U} that
give not even a hint of excitons, TDLDA+\textit{U} generates successfully the tightly
bound Frenkel excitons deep inside the Mott gap with good energy scales at 1-3 eV.
Considering the enormous binding energies of $\sim$6eV and the complexity of the multi-exciton
coupling, this degree of agreement is quite impressive, not to mention the good
\textbf{q}-dependent spectral weight that reflects the high quality of the exciton wave
functions.
Since the analytical structure of LDA+\textit{U} is the same as the hybrid functionals
and its equation of motion contains similar kernel to that of existing
GW-BSE~\cite{Onida2,Rohlfing} approximation, the same degree of success should be expected from
these approaches as well, so should above general physical picture of the formation of Frenkel
excitons.

On the other hand, our study also reveals a generic limitation of TDLDA+\textit{U} and all these
state-of-the-art approximations.
While TDLDA+$U$ puts the excitons in the right energy range, it is unable to reproduce the fine
structures of the experimental excitons satisfactorily.
Indeed, while three charge excitation energies in the Mott gap were found experimentally
(c.f.: Fig.~2), TDLDA+$U$ only manages to produce two distinct energies.
It turns out that the inability to split resulting peaks into the fine structures is dictated by
the analytical structure of the approximations.
This can be elucidated by reformulating rigorously the BSE into an effective Hamiltonian
via the creation operator of bare excitons $\left| i \right>$, $b^{\dagger}_{i}$:
\begin{equation}
H_{eff}=\sum_{i} \epsilon_{i}b^{\dagger}_{i}b_{i}-\sum_{i\neq j} I^{F}_{ij}b^{\dagger}_{i}b_{j}+\sum_{ij} I^{H}_{ij}b^{\dagger}_{i}b_{j}
\end{equation}
where $\epsilon_{i}$ denotes the energy of the bare excitons in Fig.~\ref{fig:fig2}(a).
In this representation, all the above physical effects are made transparent:
the hybridization due to the off-diagonal elements of $I^F$ and $I^H$, and the blue shift of the
excitonic energy via the diagonal elements of $I^H$.
This representation also proves that without terms that encapsulate inter-boson interactions,
$e.g.$ $I^{F}_{iji^\prime j^\prime}b^{\dagger}_{i}b^{\dagger}_{j}b_{j^\prime}b_{i^\prime}$,
this class of approximations is only capable of producing excitons in number equal to that of
the bare excitons ($3\cdot 2=6$ for NiO), subject to degeneracy under the point-group symmetry.
Thus, this generic limitation can only be removed via higher order kernel in BSE or new
generations of hybrid-functional kernels that explicitly include time-dependence.

This general conclusion is actually more stringent than those in the present TDDFT literature.
Without $\omega$-dependent $I$, our results demonstrate exciton formation with correct scale
of large binding energy.
Clearly, in this crudest level of approximation, the known necessity of non-adiabatic
(memory-dependent) kernel $f_\mathrm{xc}$ of TDDFT in the Kohn-Sham framework~\cite{Kim,Ulf,Neepa} would
originate solely from the spatial reduction of non-local $I_\mathrm{xc}$ into $f_\mathrm{xc}$~\cite{Tokatly}.
On the other hand, to allow any fine (``multiplets'') structure in strongly interacting systems,
a non-adiabatic kernel is absolutely necessary even without the spatial reduction.
Obviously, this is one key aspect that almost all the existing approximate functionals lack and
presents an essential and necessary step toward a proper description of local excitations in
strongly interacting systems, within all the existing theoretical frameworks.

In summary, a TDLDA+\textit{U} method is derived within TDDFT framework and implemented on the
basis of Wannier functions.
As a benchmark and an illustration, the formation of tightly bound excitons in NiO are analyzed
step-by-step.
A strong hybridization of multiple excitons is found essential to give reasonable excitation
energies with a large $\sim$6eV binding energy and the observed highly anisotropic spectral weight.
Our computationally inexpensive approach not only provides detailed insights into the formation
of Frenkel excitons, but also demonstrate an intrinsic non-adiabaticity of the interaction kernel
in allowing fine-structures in the excitation spectra of strongly interacting systems.
The lack of such essential feature in nowadays hybrid-functional and BSE methods advocates
the inclusion of explicit time-dependence in future design of approximate functionals.

This work is supported by the U.S. Department of Energy (DOE), Office of Basic Energy Science,
under Contract No. DE-AC02-98CH10886, and DOE CMSN.
H. C. Hsueh thanks National Science Council and NCTS of ROC for support, and also NCHC of ROC for
CPU time.

\bibliography{refs}

\begin{thebibliography}{36}
\expandafter\ifx\csname natexlab\endcsname\relax\def\natexlab#1{#1}\fi
\expandafter\ifx\csname bibnamefont\endcsname\relax
  \def\bibnamefont#1{#1}\fi
\expandafter\ifx\csname bibfnamefont\endcsname\relax
  \def\bibfnamefont#1{#1}\fi
\expandafter\ifx\csname citenamefont\endcsname\relax
  \def\citenamefont#1{#1}\fi
\expandafter\ifx\csname url\endcsname\relax
  \def\url#1{\texttt{#1}}\fi
\expandafter\ifx\csname urlprefix\endcsname\relax\def\urlprefix{URL }\fi
\providecommand{\bibinfo}[2]{#2}
\providecommand{\eprint}[2][]{\url{#2}}

\bibitem[{\citenamefont{Onida et~al.}(2002)}]{Onida}
\bibinfo{author}{\bibfnamefont{G.}~\bibnamefont{Onida}} \bibnamefont{et~al.},
  \bibinfo{journal}{Rev. Mod. Phys.} \textbf{\bibinfo{volume}{74}},
  \bibinfo{pages}{601} (\bibinfo{year}{2002}).

\bibitem[{\citenamefont{Hohenberg and Kohn}(1964)}]{Hohenberg}
\bibinfo{author}{\bibfnamefont{P.}~\bibnamefont{Hohenberg}} \bibnamefont{and}
  \bibinfo{author}{\bibfnamefont{W.}~\bibnamefont{Kohn}},
  \bibinfo{journal}{Phys. Rev.} \textbf{\bibinfo{volume}{136}},
  \bibinfo{pages}{B864} (\bibinfo{year}{1964}).

\bibitem[{\citenamefont{Kohn and Sham}(1965)}]{Kohn}
\bibinfo{author}{\bibfnamefont{W.}~\bibnamefont{Kohn}} \bibnamefont{and}
  \bibinfo{author}{\bibfnamefont{L.~J.} \bibnamefont{Sham}},
  \bibinfo{journal}{Phys. Rev.} \textbf{\bibinfo{volume}{140}},
  \bibinfo{pages}{A1133} (\bibinfo{year}{1965}).

\bibitem[{\citenamefont{Runge and Gross}(1984)}]{Runge}
\bibinfo{author}{\bibfnamefont{E.}~\bibnamefont{Runge}} \bibnamefont{and}
  \bibinfo{author}{\bibfnamefont{E.~K.~U.} \bibnamefont{Gross}},
  \bibinfo{journal}{Phys. Rev. Lett.} \textbf{\bibinfo{volume}{52}},
  \bibinfo{pages}{997} (\bibinfo{year}{1984}).

\bibitem[{\citenamefont{van Leeuwen}(1999)}]{Leeuwen}
\bibinfo{author}{\bibfnamefont{R.}~\bibnamefont{van Leeuwen}},
  \bibinfo{journal}{Phys. Rev. Lett.} \textbf{\bibinfo{volume}{82}},
  \bibinfo{pages}{3863} (\bibinfo{year}{1999}).

\bibitem[{\citenamefont{Petersilka et~al.}(1996)}]{Petersilka}
\bibinfo{author}{\bibfnamefont{M.}~\bibnamefont{Petersilka}}
  \bibnamefont{et~al.}, \bibinfo{journal}{Phys. Rev. Lett.}
  \textbf{\bibinfo{volume}{76}}, \bibinfo{pages}{1212} (\bibinfo{year}{1996}).

\bibitem[{\citenamefont{Vasiliev et~al.}(1999)}]{Vasiliev}
\bibinfo{author}{\bibfnamefont{I.}~\bibnamefont{Vasiliev}}
  \bibnamefont{et~al.}, \bibinfo{journal}{Phys. Rev. Lett.}
  \textbf{\bibinfo{volume}{82}}, \bibinfo{pages}{1919} (\bibinfo{year}{1999}).

\bibitem[{\citenamefont{Yabana and Bertsch}(1999)}]{Yabana}
\bibinfo{author}{\bibfnamefont{K.}~\bibnamefont{Yabana}} \bibnamefont{and}
  \bibinfo{author}{\bibfnamefont{G.~F.} \bibnamefont{Bertsch}},
  \bibinfo{journal}{Int. J. Quantum Chem.} \textbf{\bibinfo{volume}{75}},
  \bibinfo{pages}{55} (\bibinfo{year}{1999}).

\bibitem[{\citenamefont{Maddocks et~al.}(1994)}]{Needs}
\bibinfo{author}{\bibfnamefont{N.~E.} \bibnamefont{Maddocks}}
  \bibnamefont{et~al.}, \bibinfo{journal}{Europhys. Lett.}
  \textbf{\bibinfo{volume}{27}}, \bibinfo{pages}{681} (\bibinfo{year}{1994}).

\bibitem[{\citenamefont{Ku and Eguiluz}(1999)}]{Ku}
\bibinfo{author}{\bibfnamefont{W.}~\bibnamefont{Ku}} \bibnamefont{and}
  \bibinfo{author}{\bibfnamefont{A.~G.} \bibnamefont{Eguiluz}},
  \bibinfo{journal}{Phys. Rev. Lett.} \textbf{\bibinfo{volume}{82}},
  \bibinfo{pages}{2350} (\bibinfo{year}{1999}).

\bibitem[{\citenamefont{Ku et~al.}(2002{\natexlab{a}})}]{Ku3}
\bibinfo{author}{\bibfnamefont{W.}~\bibnamefont{Ku}} \bibnamefont{et~al.},
  \bibinfo{journal}{Phys. Rev. Lett.} \textbf{\bibinfo{volume}{88}},
  \bibinfo{pages}{057001} (\bibinfo{year}{2002}{\natexlab{a}}).

\bibitem[{\citenamefont{Tokatly and Pankratov}(2001)}]{Tokatly}
\bibinfo{author}{\bibfnamefont{I.~V.} \bibnamefont{Tokatly}} \bibnamefont{and}
  \bibinfo{author}{\bibfnamefont{O.}~\bibnamefont{Pankratov}},
  \bibinfo{journal}{Phys. Rev. Lett.} \textbf{\bibinfo{volume}{86}},
  \bibinfo{pages}{2078} (\bibinfo{year}{2001}).

\bibitem[{\citenamefont{Haverkort et~al.}(2007)}]{Haverkort}
\bibinfo{author}{\bibfnamefont{M.~W.} \bibnamefont{Haverkort}}
  \bibnamefont{et~al.}, \bibinfo{journal}{Phys. Rev. Lett.}
  \textbf{\bibinfo{volume}{99}}, \bibinfo{pages}{257401}
  (\bibinfo{year}{2007}).

\bibitem[{\citenamefont{Anisimov et~al.}(1991)}]{Anisimov2}
\bibinfo{author}{\bibfnamefont{V.~I.} \bibnamefont{Anisimov}}
  \bibnamefont{et~al.}, \bibinfo{journal}{Phys. Rev. B}
  \textbf{\bibinfo{volume}{44}}, \bibinfo{pages}{943} (\bibinfo{year}{1991}).

\bibitem[{\citenamefont{Shick et~al.}(1999)}]{Shick}
\bibinfo{author}{\bibfnamefont{A.~B.} \bibnamefont{Shick}}
  \bibnamefont{et~al.}, \bibinfo{journal}{Phys. Rev. B}
  \textbf{\bibinfo{volume}{60}}, \bibinfo{pages}{10763} (\bibinfo{year}{1999}).

\bibitem[{\citenamefont{Tran et~al.}(2006)}]{Tran}
\bibinfo{author}{\bibfnamefont{F.}~\bibnamefont{Tran}} \bibnamefont{et~al.},
  \bibinfo{journal}{Phys. Rev. B} \textbf{\bibinfo{volume}{74}},
  \bibinfo{pages}{155108} (\bibinfo{year}{2006}).

\bibitem[{\citenamefont{Leonov et~al.}(2004)}]{Leonov}
\bibinfo{author}{\bibfnamefont{I.}~\bibnamefont{Leonov}} \bibnamefont{et~al.},
  \bibinfo{journal}{Phys. Rev. Lett.} \textbf{\bibinfo{volume}{93}},
  \bibinfo{pages}{146404} (\bibinfo{year}{2004}).

\bibitem[{\citenamefont{Jeng et~al.}(2004)}]{Jeng}
\bibinfo{author}{\bibfnamefont{H.-T.} \bibnamefont{Jeng}} \bibnamefont{et~al.},
  \bibinfo{journal}{Phys. Rev. Lett.} \textbf{\bibinfo{volume}{93}},
  \bibinfo{pages}{156403} (\bibinfo{year}{2004}).

\bibitem[{\citenamefont{Yin et~al.}(2006)}]{Yin}
\bibinfo{author}{\bibfnamefont{W.-G.} \bibnamefont{Yin}} \bibnamefont{et~al.},
  \bibinfo{journal}{Phys. Rev. Lett.} \textbf{\bibinfo{volume}{96}},
  \bibinfo{pages}{116405} (\bibinfo{year}{2006}).

\bibitem[{\citenamefont{Volja et~al.}(2010)}]{Dmitri}
\bibinfo{author}{\bibfnamefont{D.}~\bibnamefont{Volja}} \bibnamefont{et~al.},
  \bibinfo{journal}{Europhys. Lett.} \textbf{\bibinfo{volume}{89}},
  \bibinfo{pages}{27008} (\bibinfo{year}{2010}).

\bibitem[{\citenamefont{Salpeter and Bethe}(1951)}]{Salpeter}
\bibinfo{author}{\bibfnamefont{E.~E.} \bibnamefont{Salpeter}} \bibnamefont{and}
  \bibinfo{author}{\bibfnamefont{H.~A.} \bibnamefont{Bethe}},
  \bibinfo{journal}{Phys. Rev.} \textbf{\bibinfo{volume}{84}},
  \bibinfo{pages}{1232} (\bibinfo{year}{1951}).

\bibitem[{\citenamefont{Ku et~al.}(2002{\natexlab{b}})}]{Ku2}
\bibinfo{author}{\bibfnamefont{W.}~\bibnamefont{Ku}} \bibnamefont{et~al.},
  \bibinfo{journal}{Phys. Rev. Lett.} \textbf{\bibinfo{volume}{89}},
  \bibinfo{pages}{167204} (\bibinfo{year}{2002}{\natexlab{b}}).

\bibitem[{\citenamefont{Larson et~al.}(2007)}]{Larson}
\bibinfo{author}{\bibfnamefont{B.~C.} \bibnamefont{Larson}}
  \bibnamefont{et~al.}, \bibinfo{journal}{Phys. Rev. Lett.}
  \textbf{\bibinfo{volume}{99}}, \bibinfo{pages}{026401}
  (\bibinfo{year}{2007}).

\bibitem[{\citenamefont{Paier et~al.}(2008)}]{Kresse}
\bibinfo{author}{\bibfnamefont{J.}~\bibnamefont{Paier}} \bibnamefont{et~al.},
  \bibinfo{journal}{Phys. Rev. B} \textbf{\bibinfo{volume}{78}},
  \bibinfo{pages}{121201(R)} (\bibinfo{year}{2008}).

\bibitem[{\citenamefont{Onida et~al.}(1995)}]{Onida2}
\bibinfo{author}{\bibfnamefont{G.}~\bibnamefont{Onida}} \bibnamefont{et~al.},
  \bibinfo{journal}{Phys. Rev. Lett.} \textbf{\bibinfo{volume}{75}},
  \bibinfo{pages}{818} (\bibinfo{year}{1995}).

\bibitem[{\citenamefont{Rohlfing and Louie}(1998)}]{Rohlfing}
\bibinfo{author}{\bibfnamefont{M.}~\bibnamefont{Rohlfing}} \bibnamefont{and}
  \bibinfo{author}{\bibfnamefont{S.~G.} \bibnamefont{Louie}},
  \bibinfo{journal}{Phys. Rev. Lett.} \textbf{\bibinfo{volume}{81}},
  \bibinfo{pages}{2312} (\bibinfo{year}{1998}).

\bibitem[{\citenamefont{Benedict et~al.}(1998)}]{Benedict}
\bibinfo{author}{\bibfnamefont{L.~X.} \bibnamefont{Benedict}}
  \bibnamefont{et~al.}, \bibinfo{journal}{Phys. Rev. Lett.}
  \textbf{\bibinfo{volume}{80}}, \bibinfo{pages}{4514} (\bibinfo{year}{1998}).

\bibitem[{\citenamefont{Baym and Kadanoff}(1961)}]{Baym}
\bibinfo{author}{\bibfnamefont{G.}~\bibnamefont{Baym}} \bibnamefont{and}
  \bibinfo{author}{\bibfnamefont{L.~P.} \bibnamefont{Kadanoff}},
  \bibinfo{journal}{Phys. Rev.} \textbf{\bibinfo{volume}{124}},
  \bibinfo{pages}{287} (\bibinfo{year}{1961}).

\bibitem[{\citenamefont{Soininen et~al.}(2005)}]{Soininen}
\bibinfo{author}{\bibfnamefont{J.~A.} \bibnamefont{Soininen}}
  \bibnamefont{et~al.}, \bibinfo{journal}{Phys. Rev. B}
  \textbf{\bibinfo{volume}{72}}, \bibinfo{pages}{045136}
  (\bibinfo{year}{2005}).

\bibitem[{sup()}]{supplement}
\eprint{See EPAPS Document No. for supplementary material.}

\bibitem[{\citenamefont{Schwarz et~al.}(2002)}]{Schwarz}
\bibinfo{author}{\bibfnamefont{K.}~\bibnamefont{Schwarz}} \bibnamefont{et~al.},
  \bibinfo{journal}{Comput. Phys. Commun.} \textbf{\bibinfo{volume}{147}},
  \bibinfo{pages}{71} (\bibinfo{year}{2002}).

\bibitem[{\citenamefont{M{\"{u}}ller and H{\"{u}}fner}(2008)}]{Muller}
\bibinfo{author}{\bibfnamefont{F.}~\bibnamefont{M{\"{u}}ller}}
  \bibnamefont{and}
  \bibinfo{author}{\bibfnamefont{S.}~\bibnamefont{H{\"{u}}fner}},
  \bibinfo{journal}{Phys. Rev. B} \textbf{\bibinfo{volume}{78}},
  \bibinfo{pages}{085438} (\bibinfo{year}{2008}).

\bibitem[{\citenamefont{Turkowski et~al.}(2009)}]{Ullrich}
\bibinfo{author}{\bibfnamefont{V.}~\bibnamefont{Turkowski}}
  \bibnamefont{et~al.}, \bibinfo{journal}{Phys. Rev. B}
  \textbf{\bibinfo{volume}{79}}, \bibinfo{pages}{233201}
  (\bibinfo{year}{2009}).

\bibitem[{\citenamefont{Kim and G{\"{o}}rling}(2002)}]{Kim}
\bibinfo{author}{\bibfnamefont{Y.}~\bibnamefont{Kim}} \bibnamefont{and}
  \bibinfo{author}{\bibfnamefont{A.}~\bibnamefont{G{\"{o}}rling}},
  \bibinfo{journal}{Phys. Rev. B} \textbf{\bibinfo{volume}{66}},
  \bibinfo{pages}{35114} (\bibinfo{year}{2002}).

\bibitem[{\citenamefont{Hellgren and von Barth}(2008)}]{Ulf}
\bibinfo{author}{\bibfnamefont{M.}~\bibnamefont{Hellgren}} \bibnamefont{and}
  \bibinfo{author}{\bibfnamefont{U.}~\bibnamefont{von Barth}},
  \bibinfo{journal}{Phys. Rev. B} \textbf{\bibinfo{volume}{78}},
  \bibinfo{pages}{115107} (\bibinfo{year}{2008}).

\bibitem[{\citenamefont{Maitra et~al.}(2002)}]{Neepa}
\bibinfo{author}{\bibfnamefont{N.~T.} \bibnamefont{Maitra}}
  \bibnamefont{et~al.}, \bibinfo{journal}{Phys. Rev. Lett.}
  \textbf{\bibinfo{volume}{89}}, \bibinfo{pages}{023002}
  (\bibinfo{year}{2002}).

\end{thebibliography}
\end{document}